\documentclass[twocolumn,aps,prl]{revtex4}
\usepackage{graphicx}
\usepackage{amsmath}
\usepackage{amssymb}
\usepackage{wasysym}
\usepackage{color}

\DeclareMathAlphabet\mathsfbi            {OT1}{cmss}{m}{sl}

\begin{document}

\title{Stokesian dynamics of pill-shaped Janus particles with stick and slip boundary conditions}

\author{Qiang Sun$^{a}$, Evert Klaseboer$^{b}$, Boo Cheong Khoo$^{a}$ and Derek Y.C. Chan$^{abcd*}$}

\affiliation{$^{a}$\mbox{Department of Mechanical Engineering, National University
of Singapore, Singapore 119260} \\
 $^{b}$\mbox{Institute of High Performance Computing, 1 Fusionopolis
Way, Singapore 138632}\\
 $^{c}$\mbox{Department of Mathematics and Statistics, The University
of Melbourne, Parkville, VIC 3010, Australia} \\
 $^{d}$\mbox{Faculty of Life and Social Sciences, Swinburne University
of Technology, Hawthorn, VIC 3122, Australia}}

\date{}

\begin{abstract}
We study the forces and torques experienced by pill-shaped Janus particles of different aspect ratios where half of the surface obeys the no-slip boundary condition and the other half obeys the Navier slip condition of varying slip lengths. Using a recently developed boundary integral formulation whereby the traditional singular behaviour of this approach is removed analytically, we quantify the strength of the forces and torques experienced by such particles in a uniform flow field in  the Stokes regime. Depending on the aspect ratio and the slip length, the force transverse to the flow direction can change sign. This is a novel property unique to the Janus nature of the particles.
\end{abstract}

\maketitle

\begin{center}
\begin{bf}
I. INTRODUCTION
\end{bf}
\end{center}

Janus particles - particles typically but not exclusively in the micrometer size range whose surfaces are comprised of two regions of different properties - can act like amphiphiles or surfactants and self-assemble to form scalable, exotic structures with requisite properties for applications from photonics to biomedicine.

It was recognised almost exactly 20 years ago~\cite{deGennes_1992} that Janus particles can assemble like `rigid' surfactants on the micron scale under the influence of Brownian forces that anneal them into energetically favorable structures under geometric constraints. Since then research has been mostly focussed on using spherical particles in which part of the surface is rendered hydrophobic and the remaining part hydrophilic. By controlling the degree of the hydrophobic/hydrophilic balance and the relative area fractions, different self-assembled structures can be made~\cite{Granick_2009, Jiang_2011} and the kinetics of the assembly process can also be tuned~\cite{Chen_2011}. The ability to produce particles of tailored shapes and aspect ratios~\cite{Rolland_2005, Chaudhary_2012} and the incorporation of magnetic properties to the Janus particles~\cite{Teo_2011} offers additional new opportunities.

Dynamic Janus properties arising from differential heat adsorption due to partial metallic coatings on the particles have led to the use of thermophoresis to actively drive Janus particles using an applied laser beam~\cite{Jiang_2010}. The possibility of exploiting different hydrodynamic slip conditions on Janus particles has received recent attention. However, because of the theoretical complexity of the problem, only perturbation analyses that assumes the slip length is small compared to the particle dimension or weak positional variation of the slip length over the particle surface have been attempted~\cite{Swan_2008, Willmott_2008, Willmott_2009}.

Here we consider the combined role of differential hydrodynamic slip and particle aspect ratio to obtain results that can serve as useful guides in quantifying the use of hydrodynamic flow fields to control pill-shaped particles with Janus hydrodynamic boundary conditions. We confine ourselves to the low Reynolds number regime that is appropriate for the flow conditions encountered in micro-fluidic devices. Our objective is to obtain accurate estimates of the forces and torques experienced by a Janus particle in a uniform flow field as a function of orientation and particle geometry. This provides a feasibility analysis of using Janus hydrodynamic properties as a way to dynamically segregate and orient particles.

\begin{center}
\begin{bf}
II. MODEL JANUS PILL
\end{bf}
\end{center}

\begin{figure} [hpt]    
\centering{}
\includegraphics[width=2.8 in]{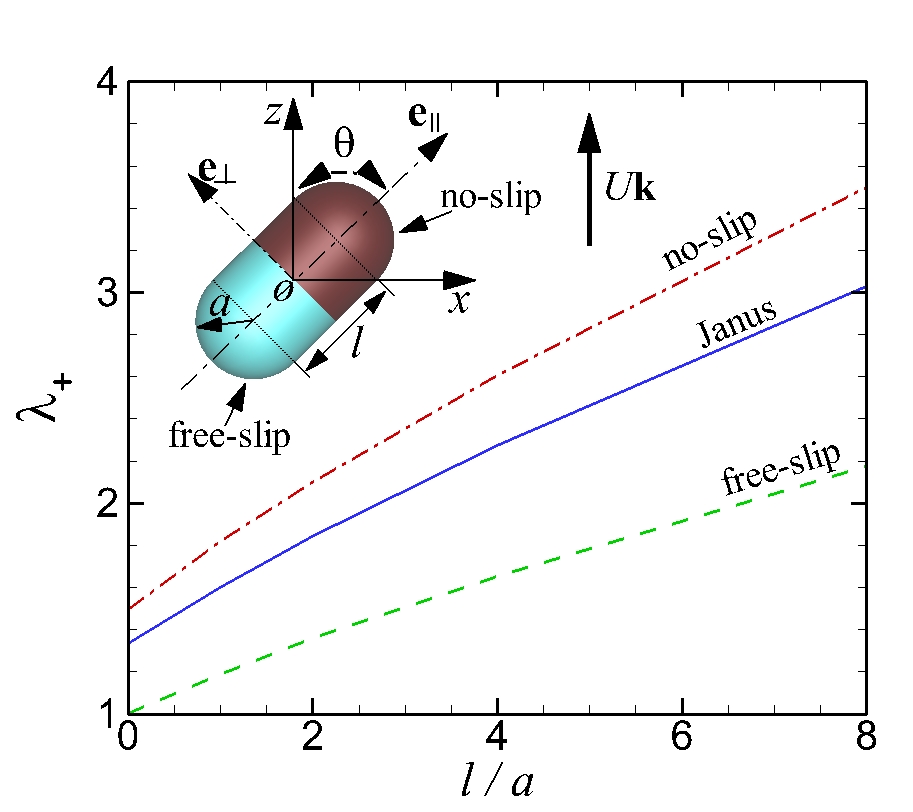}
\includegraphics[width=2.8 in]{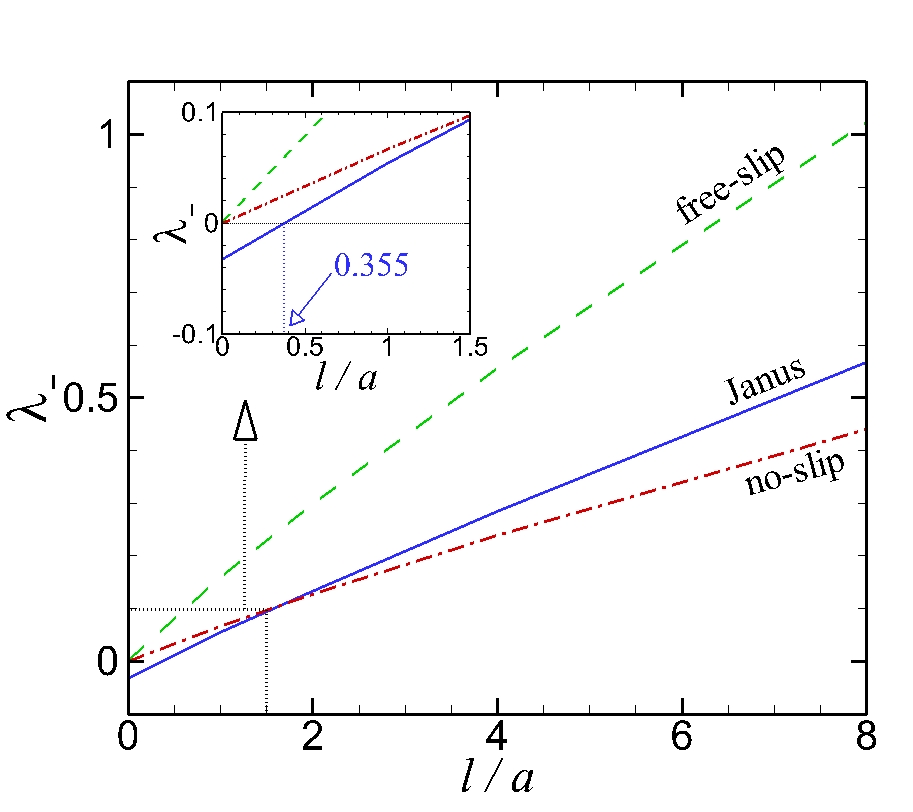}
\caption{(Color online) The dependence of the drag coefficients $\lambda_{+}$ and $\lambda_{-}$ defined in Eq.~\ref{eqn:lamdaPlusMinus} on $(l/a)$ for a Janus ({\color{blue}---}), no-slip (${\color{red}- \cdot -}$) or free-slip ({\color{green}-- --}) pills. Insets: a) The laboratory $(x, y, z)$ and body $(\mathbf{e}_{\Vert},\mathbf{e}_{\bot})$ coordinate systems. b) A close up of the region around $l/a=0.355$ in which $\lambda_{-}$ changes sign.}
\label{fig:fig1}
\end{figure}

As a prototypical model, we consider a sphero-cylinderical `Janus pill' that is made up of a right circular cylinder of radius $a$, length $l$ and capped by a hemisphere of radius $a$ at each end (Fig.~\ref{fig:fig1} inset). The boundary condition on one half of the surface along the principal body axis of the pill, $\mathbf{e}_{\Vert}$ is the familiar immobile condition with zero velocity at a solid-fluid boundary -  designated as `no-slip'. On the other half of the pill surface, we assume the surface is tangentially mobile, with a zero tangential stress condition - designated as `free-slip'. This model may be regarded as the limiting case of surfaces with finite slip lengths that we also consider briefly here. In practice, the precise experimental determination of the slip length on different types of surfaces deduced from different techniques is often complex~\cite{Rodrigues_2010, Lauga_2007, Neto_2005}.

We consider a Janus pill lying in the $xz$-plane with the body axis $\mathbf{e}_{\Vert}$ oriented at an angle $\theta$ to the $z$-direction of a uniform flow field $\mathbf{U} = U \mathbf{k}$. The novel findings are the subtle and competing effects between the aspect ratio, $l/a$ and the Janus hydrodynamic boundary condition on the force, $\mathbf{F}$ and torque, $\mathbf{T}$ that are unique consequences of the bipolar nature of the Janus particle:
\begin{itemize}
\item At a particular orientation, $\theta$, the Janus hydrodynamic boundary condition can either enhance or diminish the force $\mathbf{F}$ depending on the aspect ratio, $l/a$ (Fig. 2).
\item In particular, the component of the force, $F_x$ exerted on the Janus pill \textit{normal} to the flow direction changes sign as the aspect ratio $l/a$ increases from zero (corresponding to a Janus sphere). At $l/a = 0.355$, $F_x = 0$ (Fig. 2a).
\item The force, $F_z$ exerted on a Janus pill \textit{parallel} to the flow direction is independent of its orientation at $l/a = 0.355$ (Fig. 2b).
\item A Janus pill experiences a torque, $T_y$ about the $y$-axis that varies with its orientation and aspect ratio, $l/a \ge 0$, whereas uniform non-Janus particles experience no torque (Fig. 3).
\item Rotating a Janus pill about the $y$-axis generates a force in the equatorial plane, but not so for non-Janus pills (Fig. 4).
\end{itemize}
The above behavior that is unique to Janus particles offers opportunities to use flow fields to steer, sort, orientate and position Janus particles to create functional structures. This will be useful to enhance assembly mechanisms that only rely on energetics and geometry.

\begin{center}
\begin{bf}
III. ANALYSIS
\end{bf}
\end{center}

As the following analysis will show, the results for the forces and torques given in Figs. 2 - 4 can be characterized in terms of three functions. First, we express the external flow field $\mathbf{U} = U \mathbf{k}$ in terms of the velocity components in the body coordinate system defined by the unit vectors $\mathbf{e}_{\Vert}$ and $\mathbf{e}_{\bot}$: $\mathbf{U} = U_{\Vert}\mathbf{e}_{\Vert}+U_{\bot}\mathbf{e}_{\bot}$ with $U_{\Vert} = U\cos\theta$ and $U_{\bot} = U \sin\theta$ (Fig.~\ref{fig:fig1} inset). We define dimensionless drag coefficients: $\lambda_{\Vert}$ and $\lambda_{\bot}$ in terms of the components of the drag force on the Janus pill in the body coordinate system: $F_{\Vert}=(4\pi\mu aU_{\Vert})\,\lambda_{\Vert}$ and $F_{\bot}=(4\pi\mu aU_{\bot})\,\lambda_{\bot}$. The force components in the laboratory frame $(x,y,z)$ are then: $F_{x}(\theta, l/a) = F_{\Vert}\sin\theta-F_{\bot}\cos\theta$, $F_{z}(\theta, l/a) = F_{\Vert}\cos\theta+F_{\bot}\sin\theta$ and $F_{y}(\theta, l/a) = 0$ due to symmetry. Finally, the components of the force exerted on the Janus pill simplify to:
\begin{eqnarray}
F_{x}(\theta, l/a) &=&  (4\pi\mu aU)(\lambda_{-}\sin2\theta), \\
  \nonumber \\
F_{z}(\theta, l/a) &=&  (4\pi\mu aU)(\lambda_{+}-\lambda_{-}\cos2\theta), \\  \nonumber \\
\lambda_{+} \equiv \tfrac{1}{2}(\lambda_{\bot} &+&\lambda_{\Vert}), \quad\lambda_{-}  \equiv \tfrac{1}{2}(\lambda_{\bot} - \lambda_{\Vert}).
\label{eqn:lamdaPlusMinus}
\end{eqnarray}

The angular dependence of $F_x$ and $F_z$ are now explicit and the drag coefficients ($\lambda_{+}$, $\lambda_{-})$ depend on the aspect ratio of the Janus pill, $l/a$. Thus for a given aspect ratio, the force components $F_x$ and $F_z$ can be determined completely by calculating $F_z$, say at $\theta = 0$ and at $\theta = \tfrac{\pi}{2}$.

Linearity of Stokes flow implies that the traction on the surface of the Janus pill at orientation $\theta$ is $\mathbf{f}(\theta, l/a, U) \equiv \hat{\mathbf{f}}(\theta, l/a)\, U$, where $\hat{\mathbf{f}}(\theta, l/a)$ is the traction that corresponds to $U = 1$. The general angular dependence is
\begin{eqnarray}  \label{eqn:traction}
\mathbf{f}(\theta, l/a, U)
 &=& \hat{\mathbf{f}}(0, l/a)\,U\cos\theta+\hat{\mathbf{f}}(\tfrac{\pi}{2}, l/a)\,U\sin\theta
\end{eqnarray}

From symmetry, the torque experienced by the Janus pill only has a non-zero $y$-component given by
\begin{subequations}  \label{eqn:torque}
\begin{align}
T_{y}(\theta, l/a,U) \; \mathbf{j}
 & = \int\mathbf{r}\times\mathbf{f}(\theta, l/a, U)\; dS &\\
 & = (U\cos\theta)\int\mathbf{r}\times\mathbf{\hat{f}}(0, l/a)\; dS \quad + & \nonumber \\
 & \qquad (U\sin\theta)\int\mathbf{r}\times\mathbf{\hat{f}}(\tfrac{\pi}{2}, l/a)\; dS &\\
 & \equiv   0 \; \mathbf{j} \; \; + \;  (4\pi \mu a^2 U) \; \tau_{y}(\tfrac{\pi}{2}) \; \sin\theta \; \mathbf{j}. &
\end{align}
\end{subequations}
The first integral in Eq.~\ref{eqn:torque}b vanishes because of symmetry, hence the angular dependence of the torque is given explicitly by Eq.~\ref{eqn:torque}c and the dimensionless torque $\tau_{y}(\tfrac{\pi}{2})$ varies with the aspect ratio $(l/a)$.

\begin{center}
\begin{bf}
IV. NUMERICAL COMPUTATIONS
\end{bf}
\end{center}

The three quantities: $\lambda_{+}$, $\lambda_{-}$ and $\tau_{y}(\tfrac{\pi}{2})$ that characterize the force and torque experienced by the Janus pill under a uniform flow field in a fluid with viscosity, $\mu$, have to be computed numerically by solving the Stokes equations. A recently developed  non-singular boundary integral formulation of the solution for the Stokes equations~\cite{Klaseboer_2012} is ideally suited to handle the hydrodynamic boundary conditions that vary on different parts of the surface of the Janus pills. In this approach, the fluid velocity, $\mathbf{u}$ and the surface traction, $\mathbf{f}$ on the surface of the Janus pill are related by the following integral equation (in cartesian component form with the implicit summation convention over repeated indices)
\begin{align}  \label{eqn:BIM}
\int_{S} \left[\mu (u_i-u_{i}^{0}) - \mathsfbi{M}^{0}_{il}(x_{l}-x_{l}^{0})\right] \mathsfbi{T}_{ijk}n_{k}\text{ d}S  \qquad \\ \nonumber
 = \int_{S}\left(f_i-\mathsfbi{\Sigma}^{0}_{il}n_{l}\right)\mathsfbi{U}_{ij}\text{ d}S.
\end{align}
Here, $x_i^0$ and $x_i$ are respectively the observation and integration points on the surface $S$ with outward normals $n_i^0$ and $n_i$ and corresponding velocity, $u_i^0$ and $u_i$ and traction, $f_i^0$ and $f_i$ components, and ($r = |\mathbf{x} - \mathbf{x}_0|$)~\cite{Klaseboer_2012}
\begin{equation}
\mathsfbi{M}^{0}_{il} = f_{i}^{0}n^{0}_{l}- \tfrac{1}{4} (f_{k}^{0}n^{0}_{k})(\mathsfbi{\delta}_{il}+n_{i}^{0}n_{l}^{0}),
\end{equation}
\begin{equation}
\mathsfbi{\Sigma}^{0}_{il} =
(f_{i}^{0}n^{0}_{l}+f^{0}_{l}n_{i}^{0})-\tfrac{1}{2}(f_{k}^{0}n^{0}_{k})(\mathsfbi{\delta}_{il}+n_{i}^{0}n_{l}^{0}),
\end{equation}
\begin{equation}
\mathsfbi{U}_{ij} = \mathsfbi{\delta}_{ij} / r + \skew3\hat{x}_{i}\skew3\hat{x}_{j} / r^3, \; \;
\mathsfbi{T}_{ijk} = -6 \, \skew3\hat{x}_i\skew3\hat{x}_j\skew3\hat{x}_k / r^5.
\end{equation}

On the no-slip part of the surface, $\mathbf{u}$ is given and $\mathbf{f}$ is calculated; and on the free-slip part of the surface, the normal component of $\mathbf{u}$ and the tangential components of $\mathbf{f}$ are specified whereas the tangential components of $\mathbf{u}$ and the normal component of $\mathbf{f}$ are calculated. Once the unknown components of velocity and traction on the surface of the Janus pill are found, the force and torque on the particle can be evaluated by integrating over the surface of the pill without having to solve for the flow field in the infinite 3D domain of the fluid~\cite{Klaseboer_2012}.

As the usual singularities that occur in boundary integral methods~\cite{Becker_1992} have been removed analytically in the formulation in Eq.~\ref{eqn:BIM}, the implementation is straightforward. For example, using less than 800 nodes, our numerical method is accurate to within 1\% of the only known analytic result for the drag force on a Janus sphere~\cite{Sadhal_1983}. (The source code is available from the authors.)

\begin{figure}  
\centering{}
\includegraphics[width=2.8 in]{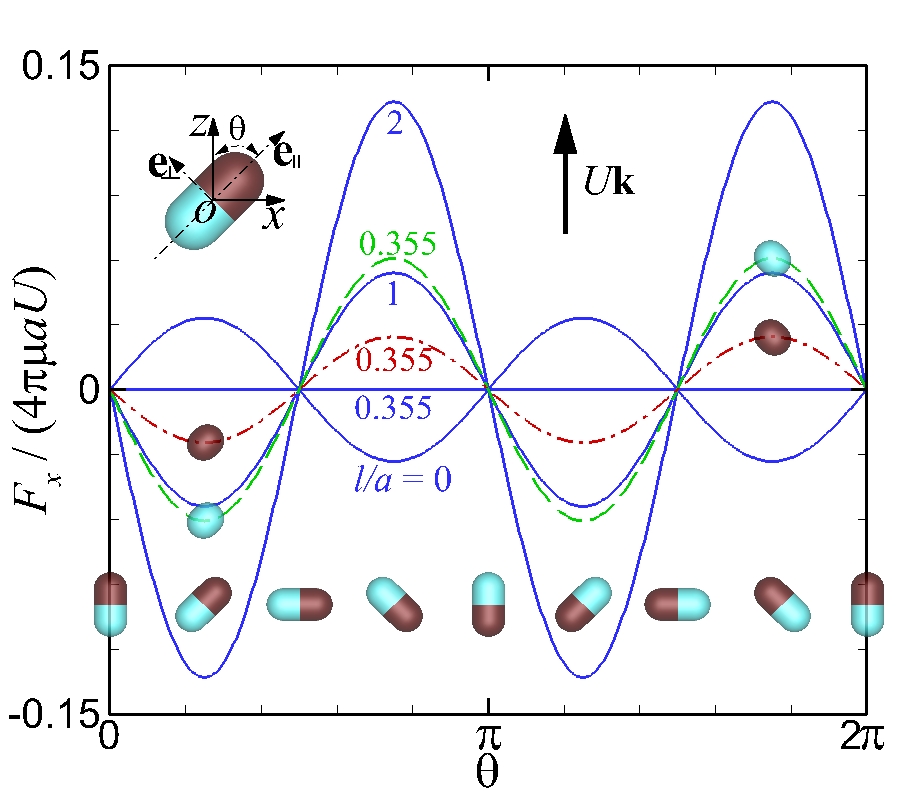}
\includegraphics[width=2.8 in]{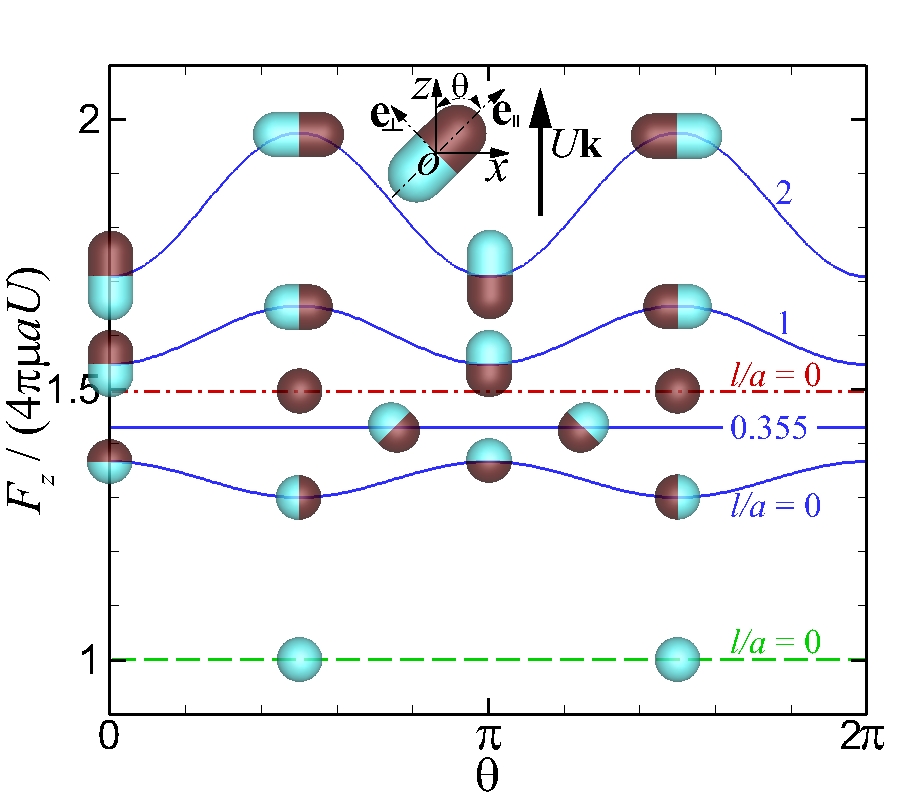}
\caption{(Color online) Force components on a Janus pill as a function of the angle of inclination, $\theta$ relative to a uniform flow in the $z$-direction for $l/a=$ 0, 0.355, 1 and 2 ({\color{blue}---}). a) The $x$-component, $F_x$ and also results for $l/a = 0.355$ for the no-slip (${\color{red}- \cdot -}$) and free-slip ({\color{green}- - -}) cases where $F_x$ for the Janus pills vanishes.  b) The $z$-component, $F_z$ and with results for $l/a=0$ that correspond to no-slip (${\color{red}- \cdot -}$) and free-slip ({\color{green}-- --}) spheres.}
\label{fig:fig2}
\end{figure}

\begin{figure}  
\centering{}
\includegraphics[width=2.8 in]{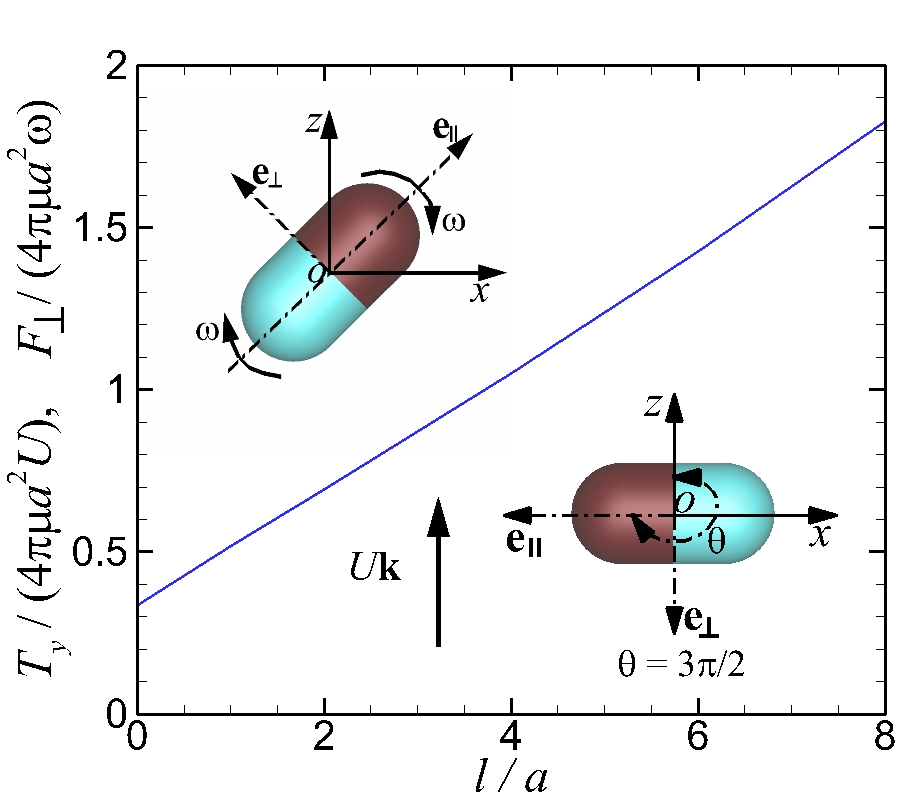}
\includegraphics[width=2.8 in]{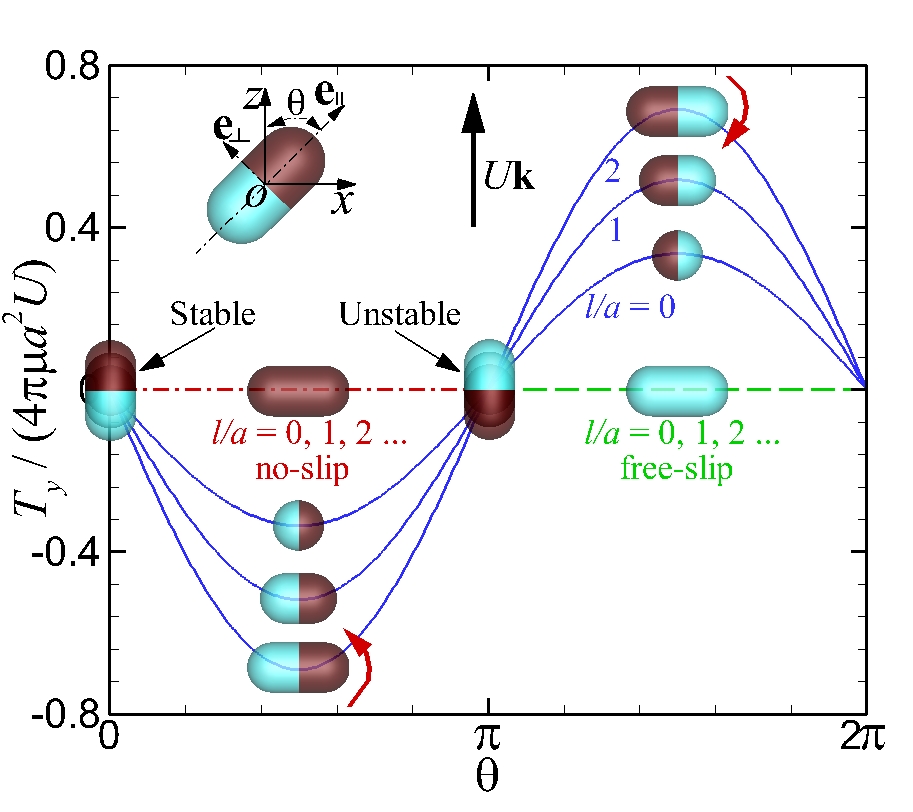}
\caption{(Color online) a) The torque, $T_y$ about the $y$-axis on a Janus pill with the body axis $\mathbf{e}_{\bot}$ oriented anti-parallel to the uniform external flow or the force, $F_{\bot}$ in the $\mathbf{e}_{\bot}$ direction of the body frame, experienced by a rotating Janus pill at angular frequency $\omega$ as a function of the aspect ratio, $l/a$. b) Variations of the torque about the $y$-axis on a Janus pill as a function of the angle of inclination, $\theta$ relative to a uniform flow in the $z$-direction for $l/a=$ 0, 1 and 2 ({\color{blue}---}). Both the uniform no-slip (${\color{red}- \cdot -}$) and free-slip ({\color{green}-- --}) pills experience no torque.}
\label{fig:fig3}
\includegraphics[width=2.8 in]{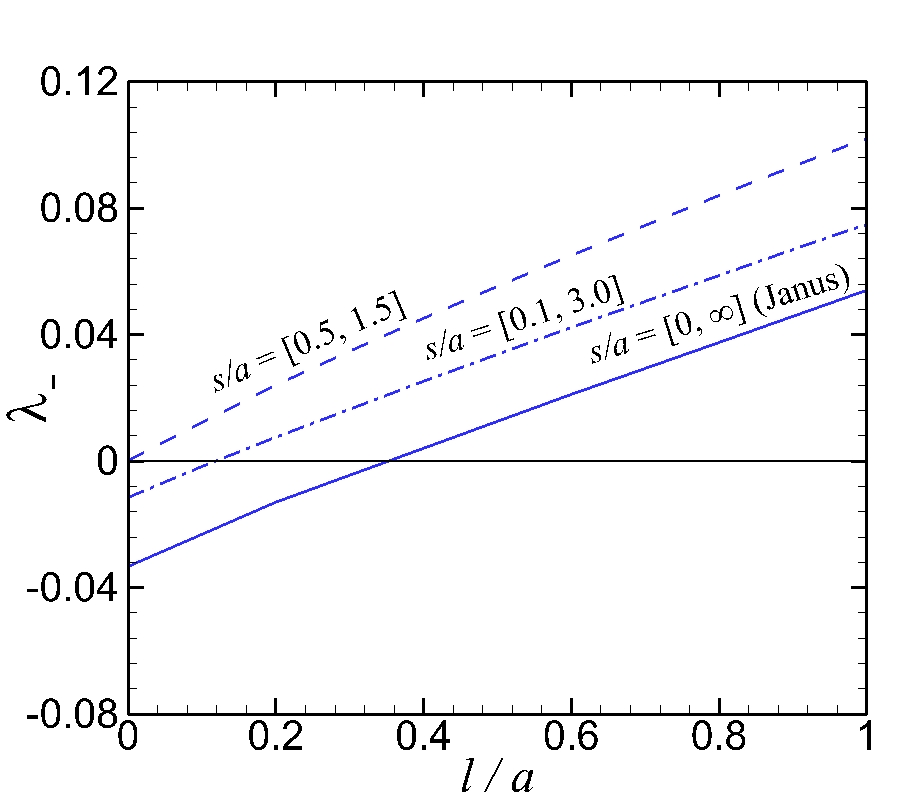}
\caption{(Color online) Variation of the drag coefficient $\lambda_{-}$ defined in Eq. 3 with aspect ratio $l/a$ for different combinations of slip lengths $s$ on each half of the Janus pill.}
\label{fig:fig4}
\end{figure}

\begin{center}
\begin{bf}
V. RESULTS
\end{bf}
\end{center}

In Fig.~1, we show the dependence of the drag coefficients $\lambda_{+}$ and $\lambda_{-}$ on the aspect ratio, $l/a$ that determine the magnitude of the force parallel, $F_z$ and perpendicular, $F_x$ to the direction of flow (Eqs. 1-3). By way of comparison, we also show the corresponding drag coefficients for a uniform no-slip and a uniform free-slip pill. The value of $\lambda_{+}$ that determines the mean value of the force in the direction of the imposed flow, $F_z$ (Eq.~2) for the Janus pill lies between the no-slip and the free-slip cases (Fig.~1a). This observation conforms with physical intuition concerning the effects of the Janus hydrodynamic boundary condition.

On the other hand, the magnitude of the force in the direction orthogonal to that of the imposed flow, $F_x$ (Eq.~1) as well as the magnitude of the periodic part of $F_z$ (Eq.~2) is determined by $\lambda_{-}$ (Fig.~1b). Here it is interesting to see that at $l/a=0.355$, $\lambda_{-}$ and hence $F_x$, the force in the direction orthogonal to that of the imposed flow, changes sign. This is a unique signature of the Janus hydrodynamic boundary condition and is an example whereby a simple flow field can be used as a dynamic mechanism to steer or sort Janus particles according to their aspect ratio~\cite{Chaudhary_2012}.

The periodic variations of the force orthogonal, $F_x$ (Fig.~2a) and parallel, $F_z$ (Fig. 2b) to the flow direction with the angle of inclination, $\theta$ of the Janus pill are illustrated explicitly (Eqs.~1-3). For comparison we also show in Fig. 2a the results for the uniform no-slip and free-slip pills at the critical ratio $l/a=0.355$ where $F_x$ for the Janus pill vanishes. Results for $F_z$ corresponding to no-slip and free-slip spheres $(l/a=0)$ are also given in Fig. 2b. Thus if a Janus particle can be aligned with an external field (eg magnetic field) with respect to the flow direction, the variation in sign and magnitude of $F_x$ can be used to steer Janus particles according to their aspect ratio, analogous to the operation of a mass spectrometer.

The force parallel to the flow direction, $F_z$, will exceed the Brownian force $(\sim kT/a)$ when $a^2 U > kT/(6 \pi \mu)$. For water under standard conditions, this condition becomes $a^2 U > 0.2$ with $a$ measured in $\mu$m and $U$ in $\mu$m/s. Since the magnitude of force in the orthogonal direction, $F_x$ that can change sign with the particle aspect ratio, is about 10\% of $F_z$ (see Fig. 2), the magnitude of $F_x$ will exceed the Brownian force when $a^2 U > 2$.

In Fig.~3a, we show the almost linear dependence of the dimensionless torque: $\tau_{y}(\tfrac{3\pi}{2})=-\tau_{y}(\tfrac{\pi}{2})$ on the aspect ratio, $l/a$ (Eq.~5). The variation of the torque with orientation to the flow field is shown in Fig.~3b. Note that for the case $l/a=0$ that corresponds to a Janus sphere, the torque is found to be
\begin{equation}
T_{y}(\theta, l/a,U) \; \mathbf{j} = -0.347 \; (4\pi \mu a^2 U) \; \sin\theta \; \mathbf{j}.
\label{eqn:torquesphere}
\end{equation}
This numerical result turns out to be within 10\% of an estimate that can be obtained by only integrating the traction over a hemisphere of a uniform no-slip sphere
\begin{equation}
T_{y}(\theta, l/a,U) \; \mathbf{j} = -\tfrac{3}{8} \; (4\pi \mu a^2 U) \; \sin\theta \; \mathbf{j}.
\label{eqn:torquesphereest}
\end{equation}

By symmetry considerations, one also expects that the Janus particle would experience a translational force if it is driven in a rotational motion about an equatorial axis, e.g. the $y$-axis. This can be achieved, for example, by the application of a time varying magnetic field on magnetic Janus particles~\cite{Teo_2011}, as in magnetic stirrers commonly encountered in chemistry laboratories. This translational force, $F_{\bot}$ is in the $\mathbf{e}_{\bot}$ direction and the variation of its magnitude with $l/a$ that has also been computed independently and is  shown in Fig.~3a. By the well-known Lorentz Reciprocity Theorem of Stokes flow that provides the linear relationship between forces or torques to linear or angular velocities~\cite{Lorentz_1907, Kim_1991, Pozrikidis_2011} the ratio $T_y/U$ and $F_{\bot}/\omega$ must be equal. In Fig.~3a we see that this is indeed the case to within numerical precision of our computation, thus providing confidence that our numerical approach and implementation are correct.

To estimate the significance of such forces, we observe that for the translational force, $F_{\bot}$ to be comparable to the Brownian force, we require $a^3 \omega > kT/(4\pi \mu)$. This inequality is readily achievable in water under standard laboratory conditions. Furthermore, the force, $F_x$ orthogonal to the flow direction, being proportional to the drag coefficient $\lambda_{-}$, (see Eq. 1 and 3) can be tuned by different combinations of the slip length, $s$ on each half of the Janus pill (Fig.~4).

\begin{center}
\begin{bf}
VI. CONCLUSIONS
\end{bf}
\end{center}

In this paper, we have demonstrated that the combination of a Janus no-slip/free-slip hydrodynamic boundary condition together with particle aspect ratio offers interesting potential for particle manipulation that includes directional sorting and alignment under uniform flow fluids as well as rotationally driven translational motion.  Although only a family of relatively simple Janus particles are considered here, the non-singular boundary element method used to find forces and torques can be easily extended to handle Janus particles of any shape and variations in slip lengths. Generalization of the present approach to other imposed flow fields such as extensional or elongational flow is straightforward. The magnitude of unique forces experienced by Janus particle exhibited here could enrich existing techniques for direct particle manipulation in, for example, the assembly of designed structures to form novel materials or in `split and mix' combinatorial library reactions on microscopic beads using flow cytometry to identify reaction products~\cite{Battersby_2002}. In a dilute suspension of Janus spheres, the viscosity increment has been studied by perturbation theory to first order in the particles volume fraction and in the ratio of the slip length to the sphere radius~\cite{Ramachandran_2009}. The boundary integral approach~\cite{Klaseboer_2012} used here can progress this study to non-spherical Janus particles with arbitrary slip properties and to include effects of hydrodynamic coupling, extending the result to higher particle volume fractions.

\textbf{Acknowledgment} This work is supported in part by an Australian Research Council Discovery Project Grant to DYCC.

$^{*}$D.Chan@unimelb.edu.au

\end{document}